\documentclass[aps,prl,twocolumn]{revtex4}
\usepackage{amsmath}
\usepackage{graphicx}

\begin{document}

\title{External Control of a Metal-Insulator Transition in (Ga,Mn)As
  Wires} 

\author{Anh Kiet Nguyen and Arne Brataas}

\affiliation{Department of Physics, Norwegian University of Science
  and Technology, NO-7491, Trondheim, Norway.}

\begin{abstract}
  Quantum transport in disordered ferromagnetic (III,Mn)V
  semiconductors is studied theoretically. Mesoscopic wires exhibit an
  Anderson disorder-induced metal-insulator transition that can be
  controlled by a weak external magnetic field. This metal-insulator
  transition should also occur in other materials with large
  anisotropic magneto resistance effects. The transition can be useful
  for studies of zero-temperature quantum critical phase transitions
  and fundamental material properties.
\end{abstract}

\maketitle




The metal-insulator transition (MIT), a zero-temperature quantum phase
transition, has since the beginning of the $20^{\text{th}}$ century
attracted much research interest~\cite{Mott:Book90,Belitz:rmp94}. The
MIT is conventionally driven by temperature, pressure, voltage,
doping, and magnetic field
amplitude~\cite{Mott:Book90,Belitz:rmp94}. Examples of systems with
MIT induced by external stress are the doped semiconductors Si:P and
Si:B~\cite{Paalanen:prl82}. Despite considerable efforts, there are
still controversies on fundamental issues such as critical exponents,
universality classes, and scaling functions for the MIT in many
systems~\cite{Mott:Book90,Belitz:rmp94,Paalanen:prl82}. Systems with
new non-destructive external knobs that continuously control the
metal-insulator transition are of great interest and can provide new
insights on zero-temperature quantum phase transitions. For example,
the determined critical exponents applies for all other systems in the
same universality class. Furthermore, deviations between theoretical
and experimental extracted critical exponents can indicate the carrier
interaction strength~\cite{Anderson:pr58}.

We show that mesoscopic wires with a large anisotropic magneto
resistance (AMR) effect~\cite{AMR} can undergo an Anderson
metal-insulator transition driven by the magnetization
\emph{direction} which may be controlled by a weak external magnetic
field.  This is a new way to externally control a MIT. In systems with
large AMR, the elastic mean free path ($l$) strongly depends on the
magnetization direction ($\theta $).  Since the localization length
($\xi $) depends on $l$ \cite{Thouless:prl77}, it also exhibits a
strong magnetization direction dependence. Additionally, quantum
confinement can enhance the $\theta$ dependence of $\xi$. In
mesoscopic quantum wires with large AMR, there can be a magnetization
direction ($\theta_{i}$) where the localization length is shorter than
the phase coherence length ($L_{\phi }$) leading to insulating
behavior. For another magnetization direction ($\theta_{m}$), the
localization length can be longer than the phase coherence length
$L_{\phi }$ giving a metallic behavior. Between those two
magnetization directions there is an \emph{Anderson disorder-induced,
  metal-insulator transition~}\cite{Anderson:pr58}.

Let us crudely estimate the required material parameters for observing
the AMR-driven MIT. For diffusive quantum wires, the conductance is $G
\!  \simeq \!  G_{\text{Sh}} l/(l+L_y)$ where $G_{\text{Sh}}$ is the
Sharvin conductance and $L_y$ is the wire
length~\cite{Thouless:prl77}. The wire becomes localized when $G \!
\sim \! e^2/h$, thus $\xi \! \sim \! l G_{\text{Sh}}
h/e^2$~\cite{Thouless:prl77}. Consequently, the ratio between the
localization lengths for two different magnetization directions is
$\xi(\theta_{i})/\xi(\theta_{m}) \! \sim \!  G_{\text{Sh}}(\theta_{i})
l(\theta_{i})/G_\text{Sh}(\theta_{m}) l(\theta_{m})$. Now, consider a
quantum wire long enough so that transport is localized for all
magnetization directions. Here, $R(\theta) \! \sim \!
e^{L_y/\xi(\theta)}$ so that the AMR resistance ratio becomes
$R(\theta_{i})/R(\theta_{m}) \! \sim \!  e^{L_y(\xi_m \!  -
  \xi_i)/(\xi_m \xi_i)}$, where $\xi_i \! = \!  \xi(\theta_{i})$. The
maximum AMR ratio is achieved when $L_y$ approaches the longest
localization length $\xi_m$ making the wire metallic for $\theta_m$
while still localized for $\theta_i$. In this case, roughly,
$R(\theta_{i})/R(\theta_{m}) \! \sim \!  e^{\xi_m/\xi_i - 1}$. Assume
that a resistance ratio around 5 is sufficient for observing the MIT
experimentally. This requires a diffusive AMR ratio
$l(\theta_{i})/l(\theta_{m}) \! \sim \!  0.6$ and a Sharvin AMR ratio
$G_\text{Sh}(\theta_{i})/G_\text{Sh}(\theta_{m}) \!  \sim \!
0.6$. Therefore, ferromagnetic wires close to the insulating phase
with AMR effects around $40\%$ or larger are good candidates for
observing the MIT at low temperatures. It is easy to experimentally
demonstrate that such a colossal magnetoresistance effect arises from
the MIT because of its extreme temperature sensitivity. The
temperature dependence of resistance and conductance fluctuations may
also be used to extract the phase coherence length and its temperature
dependence.

This low temperature metal-insulator transition will not occur in the
conventional ferromagnets Fe, Ni, and Co. The spin-orbit interaction
is weak and the AMR effect is only a couple of percent. Furthermore,
Fe, Ni, and Co are good metals, $k_{F}l\gg 1$, and the Fermi
wavelength ($2\pi/k_F$) is on the sub-nanometer scale so even
nano-scale confinement will not drive the system to the insulating
phase.  Novel ferromagnetic materials are required.

We use mesoscopic wires of ferromagnetic (III,Mn)V semiconductors
(DMS) to demonstrate the idea. Four features make DMS suitable: 1) The
mean free path strongly depends on the magnetization
direction~\cite{AMR}. 2) State-of-the art ferromagnetic (III,Mn)V
semiconductors are dirty and poor metals with $k_{F}l\!\sim \!1$
\cite{Moca:arXiv07}, even bulk DMS are close to the insulating state.
3) The Fermi wavelength is very long ($\sim \!\! 5nm$) enabling the
production of quantum wires with few transverse modes for rather
modest constriction sizes. 4) The charge carrier concentration can be
controlled by a gate potential. This may be used to decrease the Fermi
energy ($E_F$) and increase the Fermi wavelength, i.e. reducing
$k_{F}l$ and enhancing effects of quantum confinement. Both effects
drive the system closer to the insulating phase.

Localization requires the localization length to be significantly
shorter than the phase coherence length which has been measured in
(Ga,Mn)As and (In,Mn)As to be $\! \sim \!(50, 100, 200)nm$ at
temperatures $\sim \! (2000, 100, 20)mK$,
respectively~\cite{Wagner:prl06}.  We show below that for clean, $k_F
l \! \! \sim \! 10$, (Ga,Mn)As wires clear signatures of the
metal-insulator transition are observed for $L_\phi \! \sim \! 400
nm$. Typical ferromagnetic (III,Mn)V semiconductors are much dirtier,
$k_F l \! \sim \! 1$, than our simulations~\cite{Clean}. In these
samples, $L_\phi \! \sim \!  50nm$ should be sufficient to observe the
reported metal-insulator transition. The MIT we predict should,
therefore, be observable with state-of-the art DMS. Scaling relations
for a critical MIT gave good fits to measured resistivities as
functions of temperature and magnetic field in
DMS~\cite{Moca:arXiv07}. This further support the existence of a MIT
in ferromagnetic (III,Mn)V semiconductors.

Critical phase transitions occur only in the thermodynamic limit. In
finite systems, when the coherence length of critical fluctuations
approaches the system size, finite size effects set in and prevent
further development of singular behaviors~\cite{Privman:book90}. In
this sense, the metal-insulator transition we report is a
\emph{cross-over}. However, the phase coherence length and the
localization length may be externally controlled by temperature and a
gate potential. Thus, for a given set of material parameters, one may
tune $L_\phi$ and $\xi$ such that the condition $\xi_{\theta_i} \ll
L_\phi \leq \xi_{\theta_m}$ is satisfied and a MIT may be studied
within the finite size limitation~\cite{Privman:book90}.

Magnetization controlled metal-insulator switching in (Ga,Mn)As tunnel
junctions have recently been observed in
experiments~\cite{Pappert:prl06}.  The switching was interpreted to
result from the magnetization-direction dependent hole-bound-states in
a thin depleted (Ga,Mn)As layer in contact with a tunnel junction or a
constriction~\cite{Pappert:prl06}. It is easy to distinguish between
such a scenario~\cite{Pappert:prl06} and the MIT reported here using
the conductance fluctuations which for a switching
scenario~~\cite{Pappert:prl06} should be dominated by conventional
bulk conductance fluctuations. In contrast, we report an Anderson
disorder-induced quantum critical metal-insulator transition with the
corresponding quantum critical fluctuations which strongly depends on
the magnetization direction.

We capture the essential quantum transport properties of
ferromagnetic (III,Mn)V semiconductors with a discrete Hamiltonian
\begin{equation}
  H = (\gamma_1 + \frac{5}{2} \gamma_2) \frac{p^2}{2 m_e}
  - \frac{\gamma_2}{m_e} (\mathbf{p} \cdot \mathbf{J})^2 
  + \mathbf{h} \cdot \mathbf{J} + V(\mathbf{r}) \, .  
\label{Hamiltonian}
\end{equation}
Here, $\gamma_1$ and $\gamma_2$ are the Luttinger
parameters~\cite{Vurgaftman:jap01}, $m_e$ is the electron mass,
$\mathbf{J}$ is a vector of $4 \! \times \! 4$ spin matrices for $J
\!\! = \!\! 3/2$ spins, and $\mathbf{p}$ is the momentum operator in a
finite difference form, e.g. $p_x^2 f(x_i) = -\hbar^2 [f(x_{i+1}) -
2f(x_i) + f(x_{i-1})]/a_x^2$, where $a_x$ is the lattice constant
along $x$-axis. The first two terms in Eq.\eqref{Hamiltonian} is the
$4 \!  \times \! 4$ Luttinger Hamiltonian for zincblende
semiconductors in the spherical
approximation~\cite{Luttinger:pr56}. The third term in
Eq.\eqref{Hamiltonian} describes the exchange interaction between
itinerant holes and localized magnetic Mn dopants which are modeled by
a mean-field, homogeneous exchange field $ \mathbf{h} \! = \! J_{pd}
N_{\scriptscriptstyle Mn} \mathbf{M}$, where $ J_{pd}$,
$N_{\scriptscriptstyle Mn}$ and $\mathbf{M}$ are exchange interaction,
volume density and average magnetic quantum number of the local Mn
magnetic moments~\cite{FZS}, respectively. A slightly more complicated
6 band version of Eq.\eqref{Hamiltonian} quantitatively explains many
features of DMS~\cite{FZS}.


Disorder is important in ferromagnetic (III,Mn)V semiconductors, but
detailed knowledge of the impurity states and how they affect the
transport properties are lacking~ \cite{Timm:jpcm03}. The impurity
types and configurations even depend on the annealing
protocols~\cite{Timm:jpcm03}. We believe our band model with a short
range Coulomb disorder potential provides a good starting point for
theoretical studies of disorder effects on transport in ferromagnetic
(III,Mn)V semiconductors. We use Anderson impurities,
$V(\mathbf{r})=\sum_{i}V_{i}\delta_{\mathbf{r}-\mathbf{R}_{i}}$, where
$V_{i}$ and $ \mathbf{R}_{i}$ are the strength and the position of
impurity number $i$ and $\delta $ is the Kronecker delta. There is one
impurity at each lattice site. The impurity strengths are randomly and
uniformly distributed between $ -V_{0}/2$ and $V_{0}/2$.

We consider the low temperature linear response transport regime. The
conductance is calculated using the Landauer-B{\" u}tikker formula $G
= \sum_{n,m} |t_{n,m}|^2 e^2/h$,
where $t_{n,m}$ is the transmission amplitude from transverse mode $m$
to mode $n$ at $E_F$. Due to quantum coherence, our system is not
self-averaging. To study general features, we ensemble average over $
N_I \!\! = \!\! 640$ independent impurity configurations,
e.g. $\langle G \rangle = \sum_{i=1}^{N_I} G_i/N_I$. The conductance
fluctuation is defined as $\delta G = \sqrt{\langle G^2 \rangle -
  \langle G \rangle^2}$ and the impurity-averaged resistance $\langle
R \rangle = 1 / \langle G \rangle$.

Our system is a discrete, disordered conductor sandwiched between two
clean reservoirs with the same cross section. The transport direction
is along the $y$-axis. The dimensions are $L_x \!  = \! 36nm$, $L_y \!
= \! [15-1500nm]$ and $L_z \! = \!  27nm$, comparable to recent
experimental samples~\cite{Wagner:prl06}. The spacings between the
lattice points are $a_x \!\! = \!\! a_y \!\! = \!\! a_z \!\! = \!\!
1.5nm$, much smaller than the typical Fermi wavelengths used
$\lambda_F \! \sim \!  10nm$.  The hole density is assumed to vanish
outside the wire's cross section. We calculate the conductance
numerically using a stable transfer matrix method~\cite{Usuki:prb95,
  Nguyen:prl06}.

We consider homogeneous magnetizations in the $z$-$y$ plane. The
magnetization is parallel to the $z$-axis (wire) when $\theta
\!\!=\!\!0$ ($\theta \!\!=\!\!\pi /2$). Special attention is on the
directions: $\mathbf{h}_{\perp }\!\equiv \!h_{0}\hat{z}$,
$\mathbf{h}_{\angle }\!\equiv \!h_{0}(\hat{z}+\hat{y})/\sqrt{2}$, and
$\mathbf{h}_{\parallel}\!\equiv \!h_{0}\hat{y}$. The same subscripts
are used for other quantities, e.g $\langle R_{\perp }\rangle $,
$\langle R_{\angle }\rangle $, and $\langle R_{\parallel }\rangle
$. We use the model parameters: $\gamma_{1}\!=\!7.0$,
$\gamma_{2}\!=\!2.5$, $E_{F}\!=\!0.03eV$ , $h_{0}\!=\!0.06eV$ and
$V_{0}\!=\!0.25eV$ \cite{FZS}. Mesoscopic ferromagnetic (III,Mn)V
semiconducting wires are generally expected to be partly
depleted~\cite{Depletion}. Hence, $E_{F}$ should be significantly
smaller than its bulk value $E_{F}\!\sim \!0.1eV$.  Furthermore,
$E_{F}$ may also be reduced by a gate potential. The impurity strength
$V_{0}\!=\!0.25eV$ gives $l\!\sim \!20nm$ which correspond to a clean
system, $k_{F}l\!\sim \!10$~\cite{Clean}.

%
\begin{figure}[h]
\includegraphics[scale=1.0]{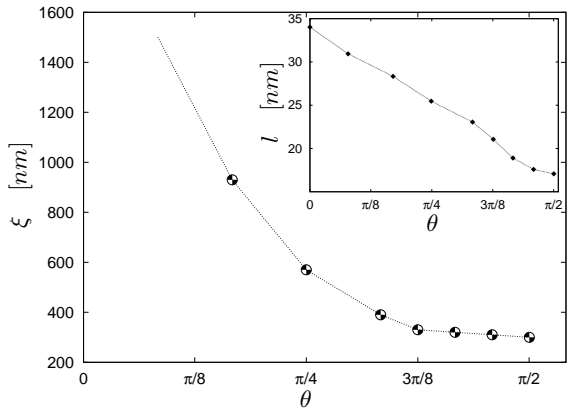} 
\caption{Localization length versus magnetization
  direction. $\protect\theta \!\!=\!\!0$ ($\protect\theta
  \!\!=\!\!\protect\pi /2$) is perpendicular (parallel) to the
  wire. Line is guide to the eye. Inset: Mean free path versus
  magnetization direction.%
  \vspace{-0.3cm}}
\label{fig:Xi_Theta}
\end{figure}
We estimate the mean free path by fitting the transmission probability
$ T=G/G_{\text{Sh}}$ to $T(L_{y})=l/(l \!+\! L_{y})$
\cite{Thouless:prl77}. Fig.~\ref{fig:Xi_Theta} inset shows the mean
free path versus magnetization direction for a (Ga,Mn)As wire. We see
that $l$ decreases from $34nm$ to $17nm$ when the magnetization
changes from perpendicular to parallel to the wire. We show below that
this magnetization direction controlled reduction of the
mean-free-path by a factor of two is sufficient to induce an
observable metal-insulator transition.

The localization length is estimated using $\langle R \rangle \!
\propto \! e^{-L_y/\xi}$ at large $L_y$. Fig.~\ref{fig:Xi_Theta} shows
that the localization length decreases strongly for increasing
magnetization angle, with a minimum $\sim \!  \! 300nm$ at $\theta
\!\! = \!\!  \pi/2$. Note that $\xi(\theta)$ decreases several times
stronger than $l(\theta)$ due to quantum confinement
~\cite{Thouless:prl77}.  Thus, for a given impurity strength there
exist a large range of wire lengths (or phase coherence lengths in the
case $L_\phi \! < \! L_y$) where $\xi(\theta \!\! = \!\! \pi/2) \ll
L_y \leq \xi(\theta \!\! = \!\! 0)$, and a metal-insulator transition
may be driven continuously by the magnetization direction.

\begin{figure}[h]
\includegraphics[scale=1.0]{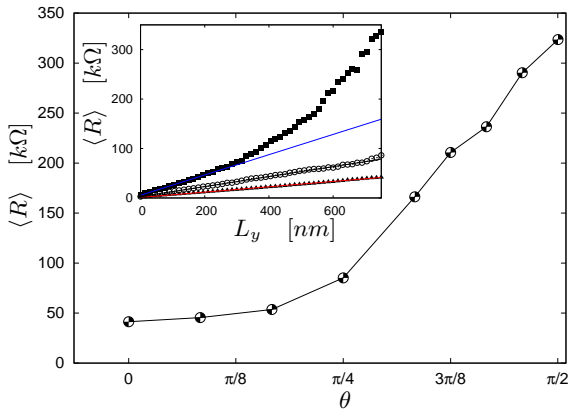} 
\caption{Impurity averaged resistance versus magnetization angle for a
  wire with length $L_{y}\!\!=\!\!750nm$. Inset: Impurity-averaged
  resistance versus wire length for $\mathbf{h}_{\perp}$ (triangles),
  $\mathbf{h}_{\protect \angle}$ (circles) and
  $\mathbf{h}_{\parallel}$ (squares). Linear lines are best fit to
  data at small $L_{y}$. %
  \vspace{-0.7cm}}
\label{fig:R_L}
\end{figure}
Fig.~\ref{fig:R_L} inset shows impurity-averaged resistances versus
wire length for the magnetizations $\mathbf{h}_{\perp }$,
$\mathbf{h}_{\angle }$ and $\mathbf{h}_{\parallel }$. The transport is
diffusive for $\mathbf{h}_{\perp }$ since $\langle R_{\perp }\rangle $
is linear for all wire lengths up to $L_{y}\!\!=\!\!750nm$. $\langle
R_{\angle }\rangle$ shows weak non-linear behavior for
$L_{y}\!>\!600nm$. For $\langle R_{\parallel }\rangle $, a clear
transition from linear to exponential behavior takes place around
$L_{y}\!\sim \!300nm$. Thus, the quantum wire is metallic for
$\mathbf{h}_{\perp }$ and localized for $ \mathbf{h}_{\parallel
}$. Between those two extreme directions, there is \emph{an Anderson
  disorder-induced metal-insulator transition which may be
  continuously controlled by the magnetization direction}.

Fig.~\ref{fig:R_L} shows the impurity averaged resistance versus
magnetization angle. We see a change in the behavior of $\langle R
\rangle$ around $\theta \!\! = \!\!  \pi/4$, after which the
resistance increases strongly for increasing angle indicating the
insulating phase. For the particular wire the resistance increases
$\sim \! \! 700\%$ when $\theta$ increases from $0$ to $\pi/2$.
Compared to our simulations, state-of-the art experimental samples are
much dirtier with much shorter localization lengths so that the AMR
should greatly increase and become a colossal anisotropy magneto
resistance effect.


Note that the anisotropy relation $\langle R_\perp \rangle \! < \!
\langle R_\parallel \rangle$ is tightly connected to the narrow wire
geometry. In ferromagnetic (III,Mn)V semiconductor films and bulk
systems, one finds experimentally the opposite AMR relation, $\langle
R_\perp \rangle \! > \!  \langle R_\parallel \rangle$~\cite{AMR}. We
have also considered a film shaped system with $L_x \! \! = \! \!
60nm$ and $L_z \! \! = \! \! 18nm$. All other parameters are the same
as above. Here, we find $\langle R_\perp \rangle \! > \!  \langle
R_\parallel \rangle$, consistent with the experimental findings. The
explanation is as follows. Due to the strong spin-orbit interaction,
the Fermi surface for heavy holes is prolonged along the magnetization
direction forcing the distribution of transverse ($k_x,k_z$)
modes/transport channels to be highly anisotropic for $\mathbf{h} \! =
\!  \mathbf{h}_\perp$ \cite{Nguyen:prl06}. For bulk and film
geometries, this anisotropy increases the probability for back
scattering and thereby increases the resistance, $\langle R_\perp
\rangle$. On the other hand, $\mathbf{h} \! = \!
\mathbf{h}_\parallel$ gives a dense and circular symmetric
distribution of transverse ($k_x,k_z$) modes which in narrow wires
leads to more confinement, i.e. stronger 1D character and consequently
more back scattering and higher resistance, $\langle R_\parallel
\rangle$.

Finally, we use the the conductance fluctuations to confirm the
existence of a metal-insulator transition. For ballistic transport
$\delta G\!=\!0$. In the transition between ballistic and diffusive
transport $\delta G$ has a peak above its universal value at $
L_{y}\!\sim \!l$ \cite{UCF_Wire}. In the diffusive regime, $\delta G$
attains a universal value~\cite{UCF}. In the localized regime,
$(\delta G)^{2}\!\propto \!\langle G\rangle $ and, thus, decreases for
increasing wire (or phase coherence length)~\cite{UCF_Wire}.
\begin{figure}[h]
\includegraphics[scale=1.0]{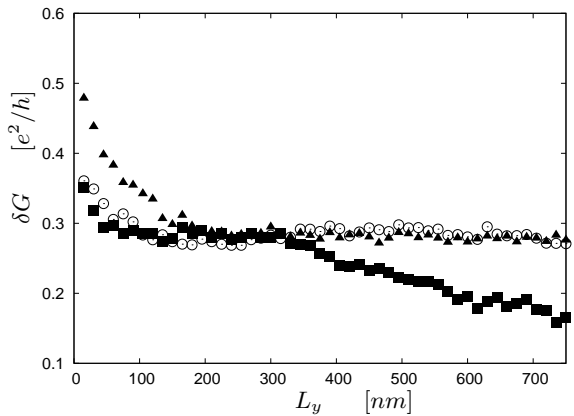} 
\caption{Conductance fluctuations versus wire length for a (Ga,Mn)As
  wire with magnetizations $\mathbf{h}_{\perp}$ (triangles),
  $\mathbf{h}_{\protect \angle}$ (circles) and
  $\mathbf{h}_{\parallel}$ (squares). %
  \vspace{-0.3cm}}
\label{fig:UCF_Length}
\end{figure}
Fig.~\ref{fig:UCF_Length} shows conductance fluctuations versus wire
length for the magnetizations $\mathbf{h}_{\perp }$,
$\mathbf{h}_{\angle }$ and $ \mathbf{h}_{\parallel }$. Since the first
shown value for $\delta G$ is at $ L_{y}\!\!=\!\!15nm$ comparable to
the mean-free-path, we do not see the rise from $\delta G\!\!=\!\!0$
expected for the $L_{y}\!\!=\!\!0$ (ballistic) case. We see that
$\delta G$ for all magnetization directions has a peak around
$L_{y}\!\sim \!l\!\sim \!20nm$ before they decrease to the same
universal value. Despite large differences in the Sharvin resistances,
conductances and mean-free-paths all magnetizations give, within
numerical uncertainties, the same universal conductance fluctuations,
$\delta G\approx 0.28e^{2}/h$. We see that the wire length must be
$\sim \!10$ times longer than the mean-free-path before the
conductance fluctuation reaches its universal value. For
$\mathbf{h}_{\parallel }$ and $L_{y}>300nm$, the conductance
fluctuation drops from its universal value and decreases for
increasing wire length. This clearly indicates a localized phase where
$\delta G\propto \sqrt{\langle G\rangle }$~\cite{UCF_Wire}, in
agreement with single parameter scaling~\cite{Abrahams:prl79}. This
drop in $\delta G$ may be used to study the phase coherence length and
its temperature dependence in the localized regime.

In conclusion, we have shown that (Ga,Mn)As mesoscopic wires of
experimental sizes exhibit at $mK$ temperatures an Anderson
disorder-induced metal-insulator transition which may be continuously
driven by the magnetization direction, which is a new way to control a
MIT. This metal-insulator transition may be useful for studies of
zero-temperature quantum critical phase transitions and fundamental
properties of ferromagnetic (III,Mn)V semiconductors. The universal
conductance fluctuations of (Ga,Mn)As wires are found to be
independent of the magnetization direction, $\delta G \! \approx \!
0.28 e^2/h$.

We thank S.\ Girvin, C. Gould, A.\ H.\ MacDonald, L.\ W.\ Molenkamp,
A.\ Sudb{\o }, Z.\ Tesanovic, C.\ Varma, and G.\ Zarand for
stimulating discussions.  This work has been supported by the Research
Council of Norway through grant no. 162742/V00, 167498/V30, computing
time through the Notur project and EC Contract IST-033749 ``DynaMax''.



\begin{thebibliography}{99}
\bibitem{Clean} State-of-the-art experimental samples are in the dirty 
limit, $k_F l \! \sim \! 1$ which should make it easier to observe the  MIT.
However, proper modelings of these systems demand, at present,  inaccessible
computing power.
\end{thebibliography}
\end{document}